\documentclass[aps,prb,twocolumn,showpacs,superscriptaddress]{revtex4-1}
\usepackage{epsfig}
\usepackage[T1]{fontenc}
\usepackage[latin9]{inputenc}
\usepackage{color}
\usepackage{amsmath, amsthm, amssymb}
\usepackage[latin9]{inputenc}
\usepackage{hyperref}
\usepackage{graphicx}
\usepackage{subfigure}

\include{mydefs}

\newcommand{\bk} { \mbox{\boldmath $k$}}
\newcommand{\bp} { \mbox{\boldmath $p$}}

\begin{document}

\title{Optical conductivity of topological Kondo insulating states}

\author{Kuang-Shing Chen}
\email{kuangshingchen@gmail.com}
\affiliation{Institut f\"ur Theoretische Physik und Astrophysik,
Universit\"at W\"urzburg, Am Hubland, D-97074 W\"urzburg, Germany}
\author{Jan Werner}
\affiliation{Institut f\"ur Theoretische Physik und Astrophysik,
Universit\"at W\"urzburg, Am Hubland, D-97074 W\"urzburg, Germany}
\author{Fakher Assaad}
\affiliation{Institut f\"ur Theoretische Physik und Astrophysik,
Universit\"at W\"urzburg, Am Hubland, D-97074 W\"urzburg, Germany}

\date{\today}

\begin{abstract}
Using real-space dynamical mean field theory with hybridization-expansion quantum Monte Carlo as a solver, we study the optical conductivity of two-dimensional topological Kondo insulating states. We consider model parameters which allow us to consider mixed valence and local moment regimes. The real space resolution inherent to our approach reveals a renormalization of the hybridization gap as one approaches the edge. Low energy transport is dominated by the helical edge state and the corresponding Drude weight scales as the coherence scale of the heavy fermion state. The concomitant renormalization of the edge state velocity leads to a constant edge local density of states. We discuss the implication of our results for the three dimensional case.
\end{abstract}


\maketitle

\section{Introduction}
The investigation of topological insulators \cite{KM_PRL1, KM_PRL2,Zhang_QSH,QSH_exp_Science1, QSH_exp_Science2} has become a very active field of research \cite{Hasan_TI_RMP,Zhang_Topo_Ins_RMP}.
Insulating in the bulk, such materials can host metallic states on the surface or at an interface. These topological surface states are protected by time-reversal symmetry, which makes them 
robust against weak disorder and interactions \cite{Luttinger_liquid_helical_edge}. The interplay between a topological band structure and correlation effects
have been studied actively \cite{Hohenad_Rev_Topo_Insulators,KMH_PRL,TBI2010,KMH2011}. In Kondo insulators, where the chemical potential is located precisely inside the band gap between the strongly renormalized bands,
a correlation induced topological ground state, the topological Kondo insulator (TKI), is proposed to be realized \cite{Coleman_PRL_TKI,Coleman_PRB_TKI}. Alongside correlation effects, the essential ingredients for the realization of topological Kondo insulating states are strong spin-orbit coupling combined with the hybridization of odd and even parity orbitals.

Kondo insulators like SmB$_6$, YbB$_{12}$ and Ce$_3$Bi$_4$Pt$_3$ exhibit a 
low-temperature resistivity which clearly deviates from an activated behavior\cite{SmB6_robustness_surface_state,SmB6_photoemission_surface_state,YbB122006,Ce3Bi4Pt3_resistivity_anomaly}. In addition, ab-initio band structure calculations performed for
SmB$_6$ \cite{Corr_Topo_Ins_Gutzwiller} have classified this material as a topological insulator, while in YbB$_{12}$ a topological crystalline insulator seems to be realized \cite{Ytterbium_Borides_TKI}.
There is experimental evidence for SmB$_6$ that the ground-state is a topological Kondo insulating state
\cite{SmB6_Topo_Exp, SmB6_Spectroscopy_Exp,SmB6_ARPES, SmB6_Surface_Hall_Exp,SmB6_gap_STS,SmB6_charge_fluctuations_PRL}, even though a direct observation of the topological surface states is still missing \cite{SmB6_origin_metallic_states}.
SmB$_6$ is a mixed valence Kondo insulator where charge fluctuations cannot be neglected and 
become apparent by a large shift of spectral weight alongside the onset of coherence \cite{SmB6_charge_fluctuations_PRL}.

Numerical studies of simple models can provide very interesting insights into pertinent questions related to topological Kondo insulators.
First, correlations can drive the system through a transition between a trivial and a non-trivial state or
between different topological states \cite{TKI_Transitions_DMFT,3D_TKI2014}. In our previous studies on the TKI model~\cite{TKI_Transitions_DMFT,JanTKI2014}, we identified an interaction-driven quantum phase transition between two distinct topological states which are connected to the $\Gamma$ and $M$ phases in the BHZ model~\cite{dislocation_pi_flux_TI}. Especially in the $\Gamma$ phase for large $U$ calculations, we found a possible non-topological Mott phase (local moment regime) featured by the divergence of the effective mass of $f$ electrons. Second, the topological properties of the system can serve as a 
convenient guide to the emergence of the coherent Fermi-liquid state \cite{JanTKI2014}. 
In Ref. \onlinecite{JanTKI2014}, we calculated the topological invariant $N_2$~\cite{Corr_Topo_Insulator_BHZ}, which is a robust measure for the topological state in the presence of correlations. 
This quantity, for different interactions $U$, shows an universal data collapse, thereby defining an energy scale $T_N$. In Ref. \onlinecite{TKI_Transitions_DMFT} it is shown that $T_N$ tracks the coherence temperature of the heavy fermion state, $T_{coh}$, and marks the dynamically induced emergence of the helical edge state. Since $T_{N}$ and $T_{coh}$ are the same scales, we will not distinguish them throughout this article.

The aim of this paper is to understand how the many-body scales show up in transport and STM experiments. We will concentrate on the optical conductivity\cite{Degiorgi_RMP1999,Rozenberg_PRB1996} and the local density of states. We study both quantities in the TKI state in the $\Gamma$ phase from the mixed valence to local moment regimes based on the real-space dynamical mean-field theory (R-DMFT)\cite{PotthoffSDDMFT} with hybridization-expansion continuous-time quantum Monte Carlo (HYB-CTQMC)\cite{PWernerPRB} as solver. We discuss the formalism in section \ref{sec:formalism}. In section \ref{sec:results} we present our results for the optical conductivity as well as for the $k$-resolved and local single-particle spectral functions. We provide a discussion and conclusion in section \ref{sec:Conclusion}.

\section{Formalism}
\label{sec:formalism}

The topological Kondo insulator (TKI)\cite{Coleman_PRB_TKI,TKI_Transitions_DMFT,JanTKI2014} is modeled by a hybridization between an odd-parity nearly localized band and an even-parity delocalized conduction band ($4f$ and $5d$ electrons in SmB$_6$ respectively) alongside a strong spin-orbit coupling. The Hamiltonian is defined by $H = H_{0}+H_{U}$ where
\begin{eqnarray}
H_{0} & = & \sum_{\bk\in BZ}\left(\begin{array}{c}
d_{\bk}^{\dagger}\\
f_{\bk}^{\dagger}
\end{array}\right)^{T}\left(\begin{array}{cc}
E_{d}(\bk) & V\mathbf{\Phi}^{\dagger}(\bk)\\
V\Phi(\bk) & E_{f}(\bk)
\end{array}\right)\left(\begin{array}{c}
d_{\bk}\\
f_{\bk}
\end{array}\right)\label{eq:H0 1}
\end{eqnarray}
and $H_{U} = U\sum_{i}f_{i\uparrow}^{\dagger}f_{i\uparrow}f_{i\downarrow}^{\dagger}f_{i\downarrow}$. Here, $\left(d_{\bk}^{\dagger},f_{\bk}^{\dagger}\right)\equiv\left(d_{\bk,\uparrow}^{\dagger},d_{\bk,\downarrow}^{\dagger},f_{\bk,\uparrow}^{\dagger},f_{\bk,\downarrow}^{\dagger}\right)$,
where the operator $d_{\bk,\sigma}^{(\dagger)}$ and $f_{\bk,\sigma}^{(\dagger)}$
annihilate (creates) the conduction electrons and the $f$ electrons
with momentum $\bk$ and pseudo-spin $\sigma$ respectively. On the
two-dimensional (2D) square lattice, we consider the dispersion $E_{d}(\bk)=-2t_d(cos(k_{x}a)+cos(k_{y}a))$
and $E_{f}(\bk)=\epsilon_{f}-2t_{f}(cos(k_{x}a)+cos(k_{y}a))$, where we take the lattice constant $a=1$. The model retains only a Kramer $f$-doublet as appropriate for rare earths with a single hole (Yb) or electron (Ce) in the $f$-shell. According to the derivation in Ref. (\onlinecite{Coleman_PRB_TKI}), the form factor $\Phi(\bk)$ contains the spin-orbit interaction and can be written as $\overrightarrow{d}(\bk)\cdot\vec{\sigma}$,
where $\overrightarrow{d}(\bk)=(2sin(k_{x}),2sin(k_{y}),0)$. To guarantee the time reversal symmetry, $\overrightarrow{d}(\bk)$ has to be an odd function of $\bk$.

Edge states are considered by using the real-space dynamical mean-field theory (R-DMFT)\cite{PotthoffSDDMFT} on a 2D square-lattice ribbon which has a periodic boundary in the $x$ direction and open boundary in the $y$ direction with $N_{y}$ layers. We obtain the layer-dependent self-energy, $\Sigma_{i}(i\omega_n)$, $i\in \left\{1\ldots N_{y}\right\}$, from the hybridization-expansion continuous-time quantum Monte Carlo (HYB-CTQMC)\cite{PWernerPRB} which is a numerical exact QMC solver and is advantageous in the strong-coupling regime. To obtain the single-particle spectra, we analytic continue the self-energy to the real-frequency axis by using the stochastic analytical continuation method\cite{KBeachMaxEnt}. We refer the reader to the appendix of Ref. \onlinecite{JanTKI2014} for a detailed description of how to  analytically continue the self energy.

At the mean-field level, the prerequisite for the TKI state is that $|\epsilon_f|>0$ and $t_f/t<0$\cite{3D_TKI2014}. Throughout this paper we focus on the $\Gamma$ phase of the TKI model from the mixed valance regime toward the local moment regime, thus we set the energy parameters $t_d=t=1$, $t_{f}=-0.2t$, $V=0.4t$, $\epsilon_{f}=-6t$ and vary the interaction $U$ from $5t$ up to $8.4t$. In this region we calculate the optical conductivity and density of states both for the ribbon and the bulk (periodic boundary conditions both in the $x$- and $y$-directions). First we will show the results using temperature $T$ as the control variable for a fixed $N_{y}=16$. Later we will consider the condition with a varied $N_{y}$.

To derive the optical conductivity for the ribbon with $N_y$ layers and
only $k_{x}$ as a good quantum number, we choose a new basis vector 
\begin{equation}
\vec{a}_{k_{x}}^{\dagger}=\left(c_{k_{x}1}^{\dagger},f_{k_{x}1}^{\dagger},c_{k_{x}2}^{\dagger},f_{k_{x}2}^{\dagger},...,c_{k_{x}N_{y}}^{\dagger},f_{k_{x}N_{y}}^{\dagger}\right)\label{eq:basis vector}
\end{equation}
which has dimension $4N_{y}$ and $c_{k_{x}i}^{\dagger}\equiv\left(c_{k_{x}i\uparrow}^{\dagger},c_{k_{x}i\downarrow}^{\dagger}\right)$
and $f_{k_{x}i}^{\dagger}\equiv\left(f_{k_{x}i\uparrow}^{\dagger},f_{k_{x}i\downarrow}^{\dagger}\right)$. The Hamiltonian (\ref{eq:H0 1}) can be rewritten as
\begin{equation}
H=\sum_{k_{x}}\vec{a}_{k_{x}}^{\dagger}\overline{H_{0}}(k_{x})\vec{a}_{k_{x}}+U\sum_{i}f_{i\uparrow}^{\dagger}f_{i\uparrow}f_{i\downarrow}^{\dagger}f_{i\downarrow},\label{eq:Hamiltonian}
\end{equation}
where $\overline{H_{0}}(k_{x})$ is a $4N_{y}\times4N_{y}$ matrix.
In this work we consider the regular part of the layer-normalized optical conductivity in the $x$ direction
\begin{equation}
Re\sigma_{xx}(\nu)=\frac{1}{N_{y}\hbar\nu}Im\left[\Lambda_{xx}(i\nu_{m}\rightarrow\hbar\nu+i\eta)\right],\label{eq:optical conductivity sigma 1-1}
\end{equation}
where $\Lambda_{xx}(i\nu_{m})$ is the current-current correlation
function and is defined as
\begin{eqnarray}
\Lambda_{xx}(i\nu_{m}) & \equiv & \int_{0}^{\beta}d\tau e^{i\nu_{m}\tau}\left\langle \overline{j_{x}}(\tau)\overline{j_{x}}(0)\right\rangle _{0}\label{eq:j-j corr tau 1}
\end{eqnarray}
with $\overline{j_{x}}(\tau)=\frac{1}{N_{k_{x}}}\sum_{k_{x}}\vec{a}_{k_{x}}^{\dagger}(\tau)\overline{J_{x}}(k_{x})\vec{a}_{k_{x}}(\tau)$
and $\overline{J_{x}}(k_{x})\equiv\frac{e}{\hbar}\frac{\partial}{\partial k_{x}}\overline{H_{0}}(k_{x})$.
Neglecting vertex corrections, we derive the layer-normalized optical conductivity
\begin{eqnarray}
Re\sigma_{xx}(\nu) & = & \frac{\pi}{N_{k_x}N_{y}}\sum_{k_{x}}\int_{-\infty}^{\infty}d\omega\frac{n_{F}(\omega+\nu)-n_{F}(\omega)}{\hbar\nu}\times\nonumber\\
 &  & Tr\left[\overline{J_{x}}(k_{x})A(k_{x},\omega+\nu)\overline{J_{x}}(k_{x})A(k_{x},\omega)\right]\label{eq:optical conductivity sigma 1}
\end{eqnarray}
with the spectral function defined as
\begin{equation}
A^{m,n}(k_{x},\omega)\equiv\frac{-1}{2\pi i}\left(G^{m,n}(k_{x},\omega^{+})-(G^{n,m}(k_{x},\omega^{+}))^{*}\right),\label{eq:spectra_A_mn}
\end{equation}
and the Green function as
\begin{equation}
G(k_{x},\omega^{+})=\left(G^{0}(k_{x},\omega^{+})^{-1}-\Sigma(\omega^{+})\right)^{-1},\label{eq:Dyson_eq}
\end{equation}
where $\omega^{+}=\omega+i0^{+}$. The unit of the optical conductivity in eq. (\ref{eq:optical conductivity sigma 1}) equals to $\frac{e^{2}}{\hbar}\approx 2.4\times10^{-4}\Omega^{-1}$. The indexes $m$ and $n$ run from $1$ to $4N_{y}$ and the right hand side of eq. (\ref{eq:Dyson_eq}) is an inversion of a $(4N_y\times4N_y)$-dimensional matrix defined under the basis of eq. (\ref{eq:basis vector}). 
Note that in the R-DMFT the self-energy $\Sigma$ is diagonal with non-zero matrix elements only on the $f$ orbitals.

\section{Results}
\label{sec:results}
\begin{figure}[tb!]
\includegraphics[width=3.1in]{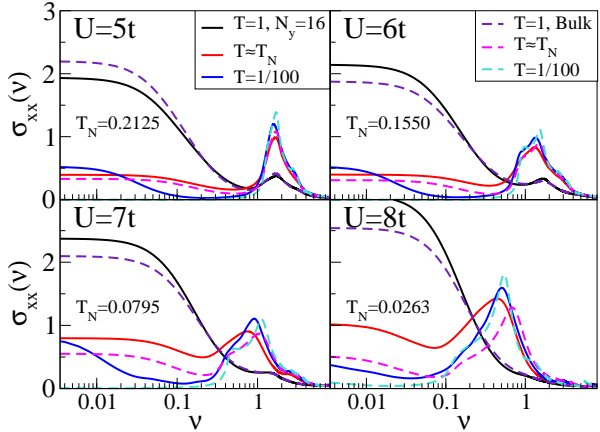}
\caption{(color online) Optical conductivity $\sigma_{xx}(\nu)$ in unit of $e^{2}/\hbar$ for $U=5t$ to $8t$ for the ribbon with $N_y=16$ (solid line) and bulk, or periodic boundary, (dashed line) topologies. We consider temperatures $T=t\gg T_{N}$, $T\approx T_{N}$ and $T=0.01t\ll T_{N}$. At $T\lesssim T_{N}$ and for the ribbon topology (solid blue line) the low-frequency finite value of $\sigma_{xx}(\nu)$ stems from the edge state.}
\label{fig:OC_U_BulkNy_}
\end{figure}

As discussed in the introduction, $T_N$\cite{TKI_Transitions_DMFT,JanTKI2014} plays the role of the coherence temperature of the heavy quasiparticles and marks the onset of the emergence of the topological edge states. In the following we will show the results with $T_N$ as a reference energy scale.  
Figure \ref{fig:OC_U_BulkNy_} shows
the optical conductivity from $U=5t$ to $8t$ both for the ribbon (open boundary in the $y$ direction with $N_{y}=16$) and the bulk calculations at three different temperatures,
$T=t\gg T_{N}$, $T\approx T_{N}$ and $T=0.01t\ll T_{N}$. When $U$
changes from $5t$ to $8t$, $T_{N}$ decreases by an order, and the
first position of the peak, the optical gap $\Delta_{opt}$ in the bulk, decreases like $\sqrt{T_{N}}$. $\Delta_{opt}$ accounts for the direct gap measured at the $\bk$ points where the non-interacting $E_d(\bk)=E_f(\bk)$. On the other hand, the Kondo hybridization gap, inversely proportional to the effective mass of $f$ electrons ($1/m_{eff}$), is estimated by the indirect gap between $\Gamma=(0,0)$ and $M=(\pi,\pi)$. We derive the $T_{N}$-scaling relations for $\Delta_{opt}$ and $1/m_{eff}$ in eq. (\ref{eq:Delta_opt}) and (\ref{eq:1_mass}) respectively. Figure \ref{fig:Delta_opt_mass} in the Appendix \ref{appendix} shows that $\Delta_{opt}\sim\sqrt{T_{N}}$ and $1/m_{eff}\sim T_{N}$. 

When $T\lesssim T_{N}$, the edge states start to develop and contribute to the low-frequency of $\sigma_{xx}(\nu)$. For $T=0.01t$, $\sigma_{xx}(\nu\rightarrow0)$ is finite for the
ribbon case (solid blue line) but decreases to zero for the bulk case (dashed cyan line), suggestive of the gapped density of states in the bulk.

\begin{figure}[tb!]
\includegraphics[width=3.1in]{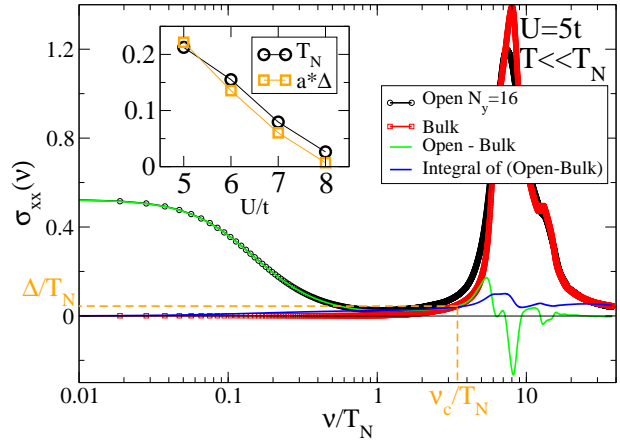}
\caption{(color online) Optical conductivity $\sigma_{xx}(\nu)$ in unit of $e^{2}/\hbar$ for $U=5t$, $T=0.01t<<T_N$ for open-boundary ribbon with $N_y=16$, bulk, and the difference between them (Open $-$ Bulk). The solid blue line shows the integral function of the difference curve. We choose the cutoff frequency $\nu_c$ roughly $2.5$. The contribution of the optical conductivity from the edge is defined as $\Delta$ in eq. (\ref{eq:Delta}). (inset) Comparison between $\Delta$ with the scaling factor $a=16$ and $T_N$ for different $U$. }
\label{fig:OC_U5B100_inset}
\end{figure}

To demonstrate that the contribution of the edge states to the optical conductivity 
scales as $T_{N}$, we consider the
integral of the difference between the open-boundary ribbon and the bulk
$\sigma_{xx}$:
\begin{equation}
\Delta\equiv\intop_{0}^{\nu_{c}}\left(\sigma_{xx}^{open}(\nu)-\sigma_{xx}^{bulk}(\nu)\right)d\nu,\label{eq:Delta}
\end{equation}
where we choose the cut-off frequency roughly at $2.5T_{N}$. Figure \ref{fig:OC_U5B100_inset}
demonstrates an example of obtaining $\Delta$ for $U=5t$ and $T=0.01t\ll T_{N}$.
The green curve is the difference curve between the open-boundary ribbon
and the bulk $\sigma_{xx}$, and the blue curve is the integral of
it. Our results show that as $\nu/T_{N}\gtrsim2.5$, the first main peak
of $\sigma_{xx}$ may start to contribute to $\Delta$ and would hence bias out result. The inset in Fig. \ref{fig:OC_U5B100_inset} shows that $\Delta\sim T_{N}$ Hence, the low-frequency optical conductivity from the topological edge state scales as the coherence temperature $T_N$.

\begin{figure}[b!]
\includegraphics[width=3in]{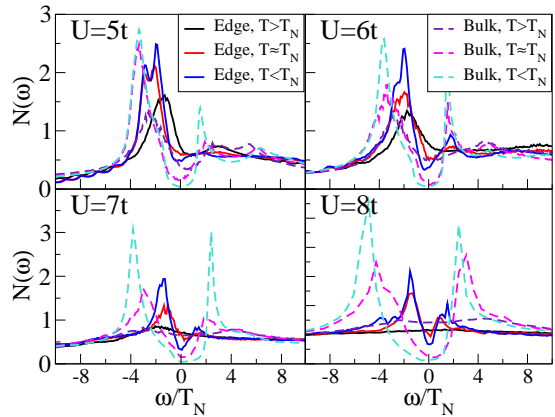}
\caption{(color online) Edge and bulk density of states from $U=5t$ to $8t$ at three temperatures, $T<T_K$, $T=T_K$, and $T>T_K$. The Kondo hybridization gap in the bulk is always larger than that at the edge at low $T$.}
\label{fig:DOS}
\end{figure}
Figure \ref{fig:DOS} shows the total ($d$ and $f$-electron) density of states at the edge and in the bulk from $U=5t$
to $8t$ at three temperature sets, $T>T_{N}$, $T\approx T_{N}$
and $T<T_{N}$. One notices that the Kondo hybridization gap in the bulk is always larger than that at edge.
One can understand this quite naturally when considering the non-interacting density
of states, $N_{0}(E_{F})$, which is smaller at the edge (1D) than in the bulk (2D). Thereby the Kondo temperature $T_{K}\propto exp(-1/JN_{0}(E_{F}))$ with $J\approx V^2/U$ \cite{book_hewson_kondo},
is smaller at the edge and the hybridization gap is also smaller. The reduction of the density of states at the surface is equally pointed out in Ref. \onlinecite{PotthoffSDDMFT}.  

\begin{figure}[t]
\includegraphics[width=3in]{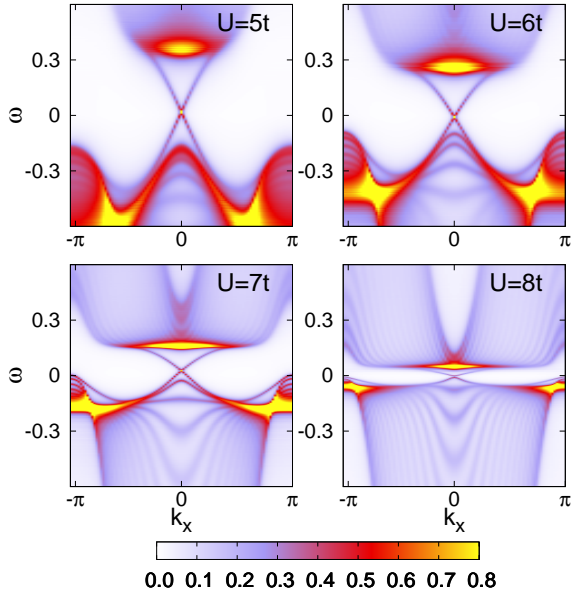}
\caption{(color online) Single-particle spectral functions, the trace of eq. (\ref{eq:spectra_A_mn}): $TrA(k_{x},\omega)$, from $U=5t$ to $8t$. The velocity of edge states, slope of the dispersion at the Fermi level, tracks the bulk gap.}
\label{fig:Akw_U_}
\end{figure}

\begin{figure}[bt]
\includegraphics[width=3in]{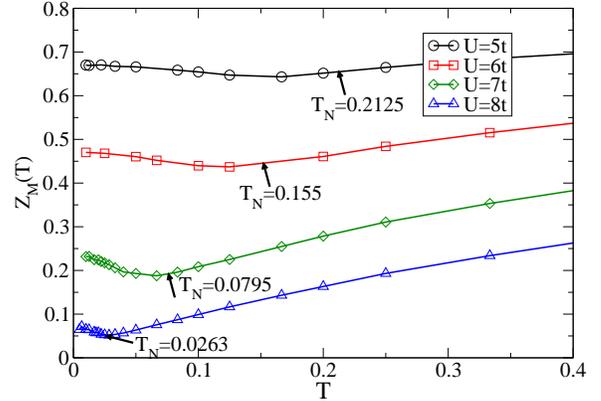}
\caption{(color online) Matsubara quasiparticle weight as a function of temperature from $U=5t$ to $8t$.  Clearly, $Z=Z_M(T\rightarrow 0)\propto T_N$.}
\label{fig:MatZ_U_}
\end{figure}

The dispersion relation at the edge near the Fermi level is linear in $k$, $\epsilon_k\approx vk$ with velocity $v\propto T_{N}$\cite{Assaad_PRB_2004}.
Figure \ref{fig:Akw_U_} clearly shows the edge state with linear dispersion in the single-particle spectral function. 
As $U$ increases from $5t$ to $8t$, the bulk gap decreases.
Note that the bulk gap tracks $T_{N}\propto T_{K}\propto exp(-U/(V^{2}N_{0}(E_{F})))$. Thus the slope of the dispersion at the Fermi level, the group velocity of the edge state $v$, also decreases and is proportional to $T_{N}$. 
One can expect that the density of states at edge near the Fermi level should be proportional to $1/v\approx 1/T_{N}$.
However, Fig. ~\ref{fig:DOS} shows that $N_{edge}(\omega=0)$ at low $T$ are roughly of the same order. 
To resolve this problem one has to include correlation effects which result in the spectral weight $Z$ of the edge state to scale as $T_{N}$. 
Here we define the Matsubara quasiparticle weight $Z_{M}$ as
\begin{equation}
Z_{M}=\left(1-\frac{Im\Sigma_{f}(i\omega_{0})}{\pi T}\right)^{-1},
\label{eq:spectral_weight}
\end{equation}
where we consider the self-energy of $f$ electrons. 
As the temperature extrapolates to zero, $Z_{M}(T\rightarrow0)\rightarrow Z$.
Figure \ref{fig:MatZ_U_} clearly shows that the spectral weight $Z$ decreases as
$U$ increases and is also proportional to $T_{N}$. 
This explains that the density of states of the edge near the Fermi level will scale as
\begin{equation}
N_{edge}(\omega\approx0)\sim Z/v\propto T_{N}/T_{N}\sim\mathcal{O}(1),\label{DOS_edge}
\end{equation}
as observed in Fig. \ref{fig:DOS}.

\begin{figure}[b]
\includegraphics[width=3in]{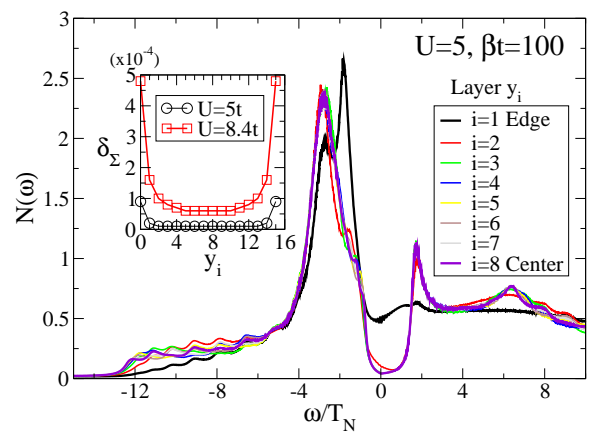}
\caption{(color online) Density of states from the edge to the center of the ribbon for $U=5t$ with $T=0.01t<<T_N$. (inset) The self-energy difference, $\delta_{\Sigma}$ in eq. (\ref{eq:delta_Sigma}), between the open-boundary ribbon with $N_y=16$ and the bulk for $U=5t$ and $8.4t$. $\delta_{\Sigma}$ illustrates that the penetration depth of the edge state is enhanced when the interaction $U$ increases.}
\label{fig:DOS_Layer_}
\end{figure}

\begin{figure}[bt]
\includegraphics[width=3in]{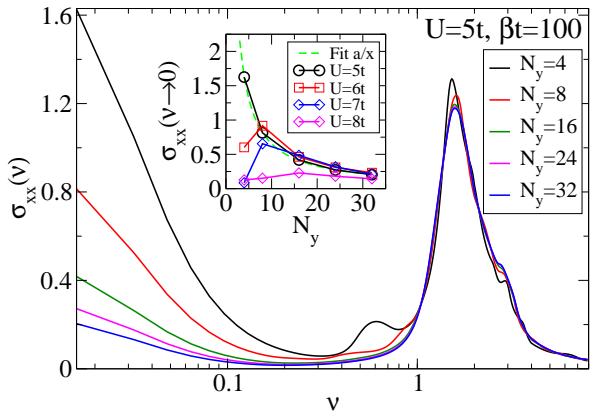}
\caption{(color online) Optical conductivity $\sigma_{xx}(\nu)$ in unit of $e^{2}/\hbar$ for $U=5t$, $T=0.01t<<T_N$ for different $N_y$. (inset) Zero-frequency of optical conductivity decays as $1/N_y$, suggestive of the contribution from the edge states coming from the first single layer.}
\label{fig:opt_cond_Layer_}
\end{figure}

Figure \ref{fig:DOS_Layer_} shows the density of states from the edge to the center of the ribbon for $U=5t$ and $T=0.01t$. Only the first layer shows significant signs of edge states and the rest of the density of states is gapped and layer-independent. To demonstrate the layer dependence of the self-energy, we define the difference between the self-energy in each layer calculated in the open-boundary ribbon with $N_y=16$ and the self-energy calculated in the bulk case:
\begin{equation}
\delta_{\Sigma}(y_i)=\sum_{\omega_{n},\sigma}\left\Vert \Sigma_{\sigma}^{open}(\omega_{n},y_i)-\Sigma_{\sigma}^{bulk}(\omega_{n})\right\Vert /(\beta\omega_{n}),\label{eq:delta_Sigma}
\end{equation}
where we divide $\beta\omega_{n}=2n+1$ to mainly consider the difference at low Matsubara frequencies and filter out and statistical error at large frequencies.

The inset of Fig. \ref{fig:DOS_Layer_} shows that $\delta_{\Sigma}(y_i)$ strongly depends on $U$, which indicates that there exists a correlation-dependent penetration depth $\xi$ for the edge state.  The notion of penetration depth allows to define bulk versus surface effects. As an example we consider the optical conductivity $\sigma_{xx}(\nu)$ in eq. (\ref{eq:optical conductivity sigma 1}). This quantity is normalized by the number of layers $N_y$, and we expect the zero-frequency $\sigma_{xx}(\nu\rightarrow 0)$ to scale as $1/N_y$ once $N_y$ is large enough, indicating that the contribution to  $\sigma_{xx}(\nu\rightarrow 0)$ from the edge state is purely a surface effect.

Figure \ref{fig:opt_cond_Layer_} shows the layer dependence of the optical conductivity $\sigma_{xx}(\nu)$ for $U=5t$ at low $T$ and the inset shows $\sigma_{xx}(\nu\rightarrow 0)$ as a function of $N_y$ for various $U$. Interestingly, $\sigma_{xx}(\nu\rightarrow 0)$ decays as $1/N_y$ when $N_y\ge8$ for $U=5t$. Once we increase $U$ up to $8t$, $\sigma_{xx}(\nu\rightarrow 0,N_y)$ starts to deviate from the fit $1/N_y$, and large values of $N_{y}$ are needed to recover the expected scaling. Hence both insets in Fig. \ref{fig:DOS_Layer_} and Fig. \ref{fig:opt_cond_Layer_} suggest that interaction effects increase the penetration depth $\xi$. 
On general grounds the  wave function of the  bulk insulating state is characterized by a localization length \cite{Kohn1964}. One expects this localization length to vary as the inverse charge gap and the penetration depth to track the localization length. Hence, one can conjecture that $\xi\sim \frac{1}{\Delta_{hyb}}\sim \frac{1}{T_{N}}$. This is consistent with our results since $T_{N}$ drops by an order of magnitude when $U$ varies from $5t$ to $8t$.

\section{Discussion and Conclusions}  
\label{sec:Conclusion}
In this article, we have used real-space DMFT to study transport and local spectroscopic properties of two dimensional TKI from the mixed valence to local moment regime. The aim of our study is to understand how many-body scales determine transport as well as the local density of states of edge states. Our first result is that the magnitude of the hybridization gap is reduced when approaching the edge. We can understand this by invoking the fact that the Kondo scale, driving the formation of the hybridization gap, scales as $T_K\approx e^{-\frac{1}{N_0(E_{F})J}}$ where $J\approx V^{2}/U$. From the expectation that the non-interacting density of state $N_{0}(E_{F})$ is reduced at the surface, follows the observation that $T_K$ as well as the hybridization gap drops when approaching the edge. The low frequency, low-T transport and local density of states are dominated by the dynamically induced helical edge states. Within the DMFT approximation, vertex corrections are neglected and it is appropriate to interpret edge transport in terms of the Drude theory of metals. Within this theory the Drude weight reads $D\approx n/m^*$ where n corresponds to the number of charge carriers and $m^*$ to their effective mass. Our result $D\approx T_N$ demonstrates that it is the heavy fermions which form the edge state. We note that $T_N$ tracks the coherence scale which is nothing but the inverse effective mass. Given this result, one can model the single-particle edge spectral function by 
\begin{equation}
A(k,\omega)=\sum_{s=\pm}Z\delta(vks-\omega)+A^{inc}(k,\omega)\label{eq:Akw_inc}
\end{equation}
with $s=\pm$ for left and right movers and $Z$ the quasiparticle residue. $A^{inc}$ accounts for the high energy spectral weight. As a consequence the single-particle density of states at the Fermi energy reads $N^{edge}(\epsilon_F)\approx Z/v$. Our numerics shows very little variation of $N^{edge}(\epsilon_F)$ from the mixed valence to local moment regimes. We thereby conclude that $Z\approx v\approx T_N$, which is confirmed by a direct claculation of single-particle spectral function. Hence, the edge quasiparticle is massless, in the sense that it obeys a massless Dirac equation, but is heavily renormalized by correlation effects since it has a small quasiparticle residue, and small velocity. The combination of small velocity and spectral weight conspire to generate a constant and scale-independent density of states.

Since the above result is based on a DMFT approach which captures fluctuations only along the imaginary time axis, one can generalize it to the three-dimensional case. Here the surface state corresponds to a two-component Dirac fermion with surface spectral function 
\begin{equation}
A(\bk ,\omega)=Z\delta(v|\bk |-\omega)+A^{inc}(\bk ,\omega),\label{eq:Akw_inc_3D}
\end{equation}
giving rise to $N(\omega)\approx Z|\omega|/v^2$ in the low frequency limit, $\omega<T_N$.
Since both $Z$ and $v$ track the coherence temperature, $T_N$, $N(\omega)\approx |\omega|/T_N$. Thereby substantial spectral weight within the hybridization gap set by $T_N$ should be visible in STM experiments\cite{SmB6_gap_STS}.

In our calculations we have omitted correlation effects beyond the DMFT approximation. For the two dimensional bulk, this provides a good description. For the corresponding one dimensional helical edge state, this is certainly not an adequate approximation since the small value of the edge state velocity is bound to render correlations beyond the DMFT approximation dominant. In particular following the work of Hohenadler et al.\cite{Luttinger_liquid_helical_edge}, we can account for spatial fluctuations along the edge. Here we can learn from previous studies and anticipate that inelastic spin-flip scattering will further reduce the spectral weight of the edge states. For the three dimensional case and corresponding two component Dirac fermion state repulsive interactions will enhance magnetism and ultimately open a mass gap by breaking time reversal symmetry\cite{TKI_DiracLiquid_arxiv,TKI_SurfaceTheory_arxiv}.

\appendix
\section{Temperature scale for the optical gap}
\label{appendix}
\begin{figure}[h]
\includegraphics[width=3in]{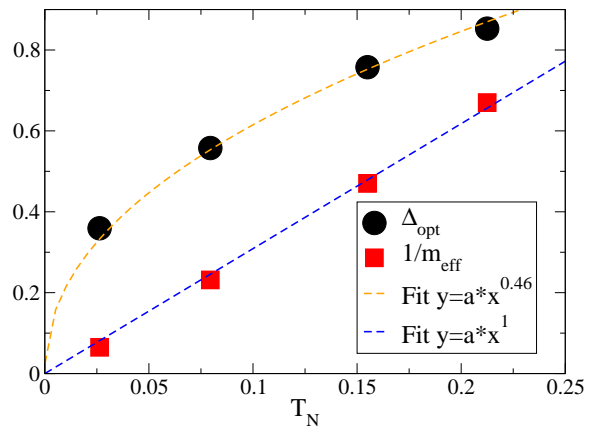}
\caption{(color online) Optical gap, $\Delta_{opt}$, and inverse of effective mass of $f$ electrons, $1/m_{eff}$, as a function of $T_N$. }
\label{fig:Delta_opt_mass}
\end{figure}

For the topological band insulator (TBI), we can rewrite the non-interacting Hamiltonian in eq. (\ref{eq:H0 1}) shifted by the chemical potential, $\tilde{H_{0}}=H_{0}-\mu$, in terms of $\Gamma$ matrices:
\begin{eqnarray}
\tilde{H_{0}}(\bk) & = & \left(\begin{array}{cc}
\tilde{\epsilon_{d}}(\bk) & (\vec{\sigma}\cdot\vec{V}(\bk))^{\dagger}\\
\vec{\sigma}\cdot\vec{V}(\bk) & \tilde{\epsilon_{f}}(\bk)
\end{array}\right)\nonumber \\
  & = & g_{0}(\bk)+\sum_{a=1}^{5}g_{a}(\bk)\Gamma_{a},
\end{eqnarray}
where 
$\tilde{\epsilon_{d}}(\bk)=E_{d}(\bk)-\mu$,
$\tilde{\epsilon_{f}}(\bk)=E_{f}(\bk)-\mu$,
$\vec{V}(\bk)=V\vec{d}(\bk)$, and
\begin{eqnarray}
g_{0}(\bk) & = & \frac{1}{2}\left(\tilde{\epsilon_{d}}(\bk)+\tilde{\epsilon_{f}}(\bk)\right)\nonumber\\
g_{1}(\bk) & = & \frac{1}{2}\left(\tilde{\epsilon_{d}}(\bk)-\tilde{\epsilon_{f}}(\bk)\right)\nonumber\\
g_{2}(\bk) & = & 0\\
(g_{3},g_{4},g_{5})(\bk) & = & (V_{x},V_{y},V_{z})(\bk).\nonumber
\end{eqnarray}
Our choice of $\Gamma$ matrices reads:
\begin{equation}
\begin{array}{ccc}
\Gamma_{1}=\sigma_{z}\otimes I_{2\times2}, & \Gamma_{2}=\sigma_{y}\otimes I_{2\times2},\\
\Gamma_{3}=\sigma_{x}\otimes\sigma_{x}, & \Gamma_{4}=\sigma_{x}\otimes\sigma_{y}, & \Gamma_{5}=\sigma_{x}\otimes\sigma_{z},
\end{array}
\end{equation}
and satisfy the relation
\begin{equation}
\left\{\Gamma_{a},\Gamma_{b}\right\} = 2\delta_{a,b}\mathit{I}_{4\times 4}.
\end{equation}
One can show that
\begin{equation}
\left(\tilde{H_{0}}(\bk)-g_{0}(\bk) \right)^{2}=\sum_{a=1}^{5}g_{a}^{2}(\bk)\Gamma_{a}^{2},
\end{equation}
such that the eigenvalues of the Hamiltonian read
\begin{eqnarray}
E^{\pm}(\bk) & = & g_{0}(\bk)\pm|\vec{g}(\bk)|\nonumber\\
       & = & \left(\tilde{\epsilon_{d}}(\bk)+\tilde{\epsilon_{f}}(\bk)\right)/2\pm\nonumber\\
       &   & \sqrt{\left[\left(\tilde{\epsilon_{d}}(\bk)-\tilde{\epsilon_{f}}(\bk)\right)/2\right]^{2}+|\vec{V}(\bk)|^{2}}.
\label{eq:Eigenvalues_H0}
\end{eqnarray}
To estimate the direct optical gap, we first set $\vec{V}(\bk)$ to zero and consider the set of $\bk$-points which satisfy $\tilde{\epsilon_{d}}(\bk)=\tilde{\epsilon_{f}}(\bk)$. We denote this set of $\bk$-points by $\bp$. 
Once $\vec{V}(\bk)\ne0$, the optical gap is a direct gap at $\bk=\bp$ and it reads
\begin{equation}
\Delta_{opt}=E^{+}(\bp)-E^{-}(\bp)=2|\vec{V}(\bp)|.
\end{equation}
On the other hand, the hybridization gap $\Delta_{hyb}$ is an indirect gap. An estimate is obtained by considering the time reversed invariant momentum, $\Gamma=(0,0)$ and $M=(\pi,\pi)$. Here we calculate $\Delta_{hyb}$ in the $\Gamma$ phase\cite{TKI_Transitions_DMFT} for $U=0$ by choosing $-4(t-t_{f})<\epsilon_{f}<0$. From the eq. (\ref{eq:Eigenvalues_H0}), we obtain
\begin{equation}
\Delta_{hyb}=E^{+}(\bk=\Gamma)-E^{-}(\bk=M)=8|t_{f}|.
\end{equation}
In the slave boson approximation\cite{Coleman_PRB_TKI}, 
\begin{eqnarray}
|\vec{V}(\bp)| & \rightarrow & b|\vec{V}(\bp)|\\
t_{f} & \rightarrow & b^{2}t_{f},
\end{eqnarray}
where the factor $b$ accounts for the band renormalization present due to the correlation effects. The coherence scale is set by $b^{2}$ which is proportional to $T_N$\cite{Assaad_PRB_2004} or the inverse of effective mass of the $f$ electrons. Thereby one can expect that 
\begin{eqnarray}
\Delta_{opt} & \approx & \sqrt{T_{N}},\label{eq:Delta_opt}\\
\Delta_{hyb}\sim 1/m_{eff} & \approx & T_{N}\label{eq:1_mass}.
\end{eqnarray}
Figure \ref{fig:Delta_opt_mass} shows the optical gap $\Delta_{opt}$ and inverse of effective mass of $f$ electrons as function of $T_N$. $\Delta_{opt}$ is measured from the half width of the bulk gap in the optical conductivity $\sigma_{xx}(\nu)$ at low $T$ in Fig. \ref{fig:OC_U_BulkNy_}. $1/m_{eff}$ is the spectral weight $Z=Z_{M}(T\rightarrow 0)$ in the eq. (\ref{eq:spectral_weight}). Basically the relations (\ref{eq:Delta_opt}) and (\ref{eq:1_mass}) hold as a function of $T_N$ for different interactions $U$.

\begin{acknowledgments}
We would like to thank M. Bercx, G. Li, C.-H. Min  as well as F.  Reinert for discussion.
Funding from the DFG under the grant number AS120/8-2 (Forschergruppe FOR 1346) is acknowledged.
We thank the  J\"ulich Supercomputing Centre  and the Leibniz-Rechenzentrum in Munich for generous allocation of CPU time.
\end{acknowledgments}

\bibliography{TKIOptCond}

\end{document}